\documentclass{elsart}
\usepackage{color,setspace,times}
\usepackage{amssymb,amsmath,graphicx,color,rotating,subfigure,url}
\usepackage[square,sort&compress,comma]{natbib}
\usepackage{lineno}

\journal{Physica A}

\begin{document}

\begin{frontmatter}

\title{Empirical shape function of limit-order books in the Chinese stock market}
\author[SB,SS]{Gao-Feng Gu},
\author[SZSE]{Wei Chen},
\author[SB,SS,CES,RCSE]{Wei-Xing Zhou\corauthref{cor}}
\corauth[cor]{Corresponding address: 130 Meilong Road, P.O. Box 114,
School of Business, East China University of Science and Technology,
Shanghai 200237, China. Tel.: +86 21 64253634; fax: +86 21
64253152.}
\ead{wxzhou@ecust.edu.cn} %

\address[SB]{School of Business, East China University of Science and Technology, Shanghai 200237, PR China}
\address[SS]{School of Science, East China University of Science and Technology, Shanghai 200237, PR China}
\address[SZSE]{Shenzhen Stock Exchange, 5045 Shennan East Road, Shenzhen 518010, PR China}
\address[CES]{Research Center for Econophysics, East China University of Science and Technology, Shanghai 200237, PR China}
\address[RCSE]{Research Center of Systems Engineering, East China University of Science and Technology, Shanghai 200237, PR China}

\begin{abstract}
We have analyzed the statistical probabilities of limit-order book
(LOB) shape through building the book using the ultra-high-frequency
data from 23 liquid stocks traded on the Shenzhen Stock Exchange in
2003. We find that the averaged LOB shape has a maximum away from
the same best price for both buy and sell LOBs. The LOB shape
function has nice exponential form in the right tail. The buy LOB is
found to be abnormally thicker for the price levels close to the
same best although there are much more sell orders on the book. We
also find that the LOB shape functions for both buy and sell sides
have periodic peaks with a period of five. The 1-min averaged
volumes at fixed tick level follow lognormal distributions, except
for the left tails which display power-law behaviors, and exhibit
long memory. Academic implications of our empirical results are also
discussed briefly.
\end{abstract}


\end{frontmatter}

\section{Introduction}
 \label{introduction}

In an order-driven market, limit-order book (LOB) is a queue of
orders waiting to be executed and it is the base of continuous
double auction mechanism. Orders in the book are sorted according to
{\em{price-time priority}}. The construction of LOB is a dynamic
process. Effective limit orders whose prices do not penetrate the
opposite best price are stored in the book, while an effective
market order with the price penetrating the opposite best
immediately causes a transaction and removes the corresponding
orders in the opposite book. In addition, cancelations can also
remove the orders in the LOB.

The price levels in the limit-order book are discrete. The
difference between two adjacent price levels is the tick size $u$.
It is 0.01 RMB for all stocks in the Chinese market. The price level
$\Delta$ at any given time $t$ can be defined as follows
\begin{equation}
\Delta = \left\{
\begin{array}{llll}
 (p_b - p)/u + 1 &&& {\rm{for~buy~orders}} \\
 (p - p_a)/u + 1 &&& {\rm{for~sell~orders}},
\end{array}
\right.
\label{Eq:x}
\end{equation}
where $p$ is an allowed price in the LOB and $p_b$ and $p_a$ are the
best bid and best ask, respectively. According to the definition,
$\Delta = 1$ stands for the position at the best bid (ask) in the
buy (sell) LOB. Denote $V_b(\Delta,t)$ (respectively
$V_s(\Delta,t)$) as the volume at level $\Delta$ in the buy
(respectively sell) LOB at event time $t$. $V_b(\Delta,t)$ and
$V_s(\Delta,t)$ can be viewed as the instant LOB shape functions on
the buy and sell sides, respectively.

The LOB shape function is of crucial importance in the research of
market microstructure theory of order-driven markets. A brief
discussion is in order. The shape of the LOB affects a trader's
strategy and thus influences order aggressiveness
\cite{Ranaldo-2004-JFinM}. Second, the LOB shape determines the
virtual price impact. The price impact $I(\omega)$ of a virtual
market order of size $\omega$ can be determined as follows
\cite{Challet-Stinchcombe-2001-PA,Maslov-Mills-2001-PA,Weber-Rosenow-2005-QF}
\begin{equation}
 I(\omega) = u\times\sup\left\{n:\sum_{\Delta=1}^{n}V(\Delta,t)\leqslant\omega\right\}~.
\end{equation}
It is found that the virtual price impact is much stronger than the
actual impact \cite{Weber-Rosenow-2005-QF} and large price
fluctuations are not necessarily caused by large orders but rather
the liquidity
\cite{Farmer-Gillemot-Lillo-Mike-Sen-2004-QF,Weber-Rosenow-2006-QF}.
It is rational that a large trader prefers to split his large order
and submit when the opposite LOB is thick such that the price does
not change much. In contrast, an impatient small trader might submit
an small order when the opposite LOB is thin for small $\Delta$'s,
since usually he does not have ensuing orders. The optimal trading
strategy of a large order also depends on the average LOB shape
\cite{Obizhaeva-Wang-2008-JFinM,Alfonsi-Schied-Schulz-2007-xxx},
which could be improved if one considers the instant LOB shape
function rather than the average.

When we want to investigate the aforementioned topics analytically,
the LOB shape function is usually treated as continuous. In the
derivation of an optimal execution strategy, many unrealistic LOB
shape functions have been proposed
\cite{Obizhaeva-Wang-2008-JFinM,Alfonsi-Schied-Schulz-2007-xxx}.
This makes the framework less useful in practice and calls for a
realistic shape function. Indeed, the empirical LOB shape function
has been investigated in different stock markets. Bouchaud {\em et
al}. found that the LOB shape of individual liquid stocks on the
Paris Bourse (February 2001) is symmetrical for buys and sells and
has a maximum away from the current bid (ask)
\cite{Bouchaud-Mezard-Potters-2002-QF}. They also found that the
distribution of order size at the bid (or ask) can be fitted by a
gamma distribution \cite{Bouchaud-Mezard-Potters-2002-QF}. Potters
and Bouchaud investigated three stocks traded on the Nasdaq Stock
Market and found that all the LOB shape functions are buy/sell
symmetric and only one stock reaches a maximum before relaxation
\cite{Potters-Bouchaud-2003-PA}. Similar results on the shape
function are also reported using other market data
\cite{Challet-Stinchcombe-2001-PA,Maslov-Mills-2001-PA,Weber-Rosenow-2005-QF,Eisler-Kertesz-Lillo-2007-PSPIE}.

In this paper, we shall study in detail the LOB shape of 23 liquid
stocks traded on the Shenzhen Stock Exchange (SZSE) in China. The
rest of the paper is organized as follows. In
Section~\ref{database}, we describe briefly the database we adopt.
Section~\ref{averageshape} introduces the average shape of buy and
sell LOBs. We then discuss in Section~\ref{sp} the probability
distributions and time dependency of volumes at the first three
best. The last section concludes.

\section{Data sets}
\label{database}

The Chinese stock market is a pure order-driven market where orders
are matched resulting in transactions. Our data contain
ultra-high-frequency data of 23 liquid stocks listed on the Shenzhen
Stock Exchange in 2003 \cite{Gu-Chen-Zhou-2008a-PA}. We find that
the results for different stocks are qualitatively similar. Hence we
will present the results for a very liquid stock. In 2003, only
limit orders were allowed to submit and the market constituted
opening call auction, cooling period and continuous double auction.
We focus on the LOB in continuous double auction.

As an example, our presentation is based on the order flow data for
a stock named Shenzhen Development Bank Co., LTD (code 000001),
whose time stamps are accurate to 0.01 second including details of
every event, with the information containing date, order size, limit
price, time, best bid, best ask, transaction volume, and
aggressiveness identifier (which identifies whether a record is a
buy order, a sell order, or a cancelation). The database totally
records $3,925,832$ events, including $1,718,156$ buy orders,
$1,595,961$ sell orders, $598,750$ cancelations and $12,965$ invalid
orders. Using this nice database, we rebuild the LOB according to
the trading rules \cite{Gu-Chen-Zhou-2007-EPJB} and study the
statistical probabilities of LOB shape.

\section{Averaged shape}
 \label{averageshape}

In the continuous double auction mechanism, order placement adds
volume to the book, while order cancelation or transaction removes
volume from the book. It is clear that these three types of events
(order placement, order cancelation and transaction) can change the
shape of the LOB. In what follows we use event time, not clock time.
In this way, the event time $t$ advances by 1 when an event occurs.
At every time $t$, we have an instant LOB shape $V_{b,s}(\Delta,t)$
on each side (buy or sell). The averaged shape of the buy (sell) LOB
can be calculated as follows
\begin{equation}
V_{b,s}(\Delta) =
 \frac{1}{M}\sum_{t = 1}^M V_{b,s}(\Delta,t)~,
\label{Eq:V}
\end{equation}
where $M$ is the number of total events in 2003 for the stock we
analyzed.

\begin{figure}[htb]
\centering
\includegraphics[width=6.5cm]{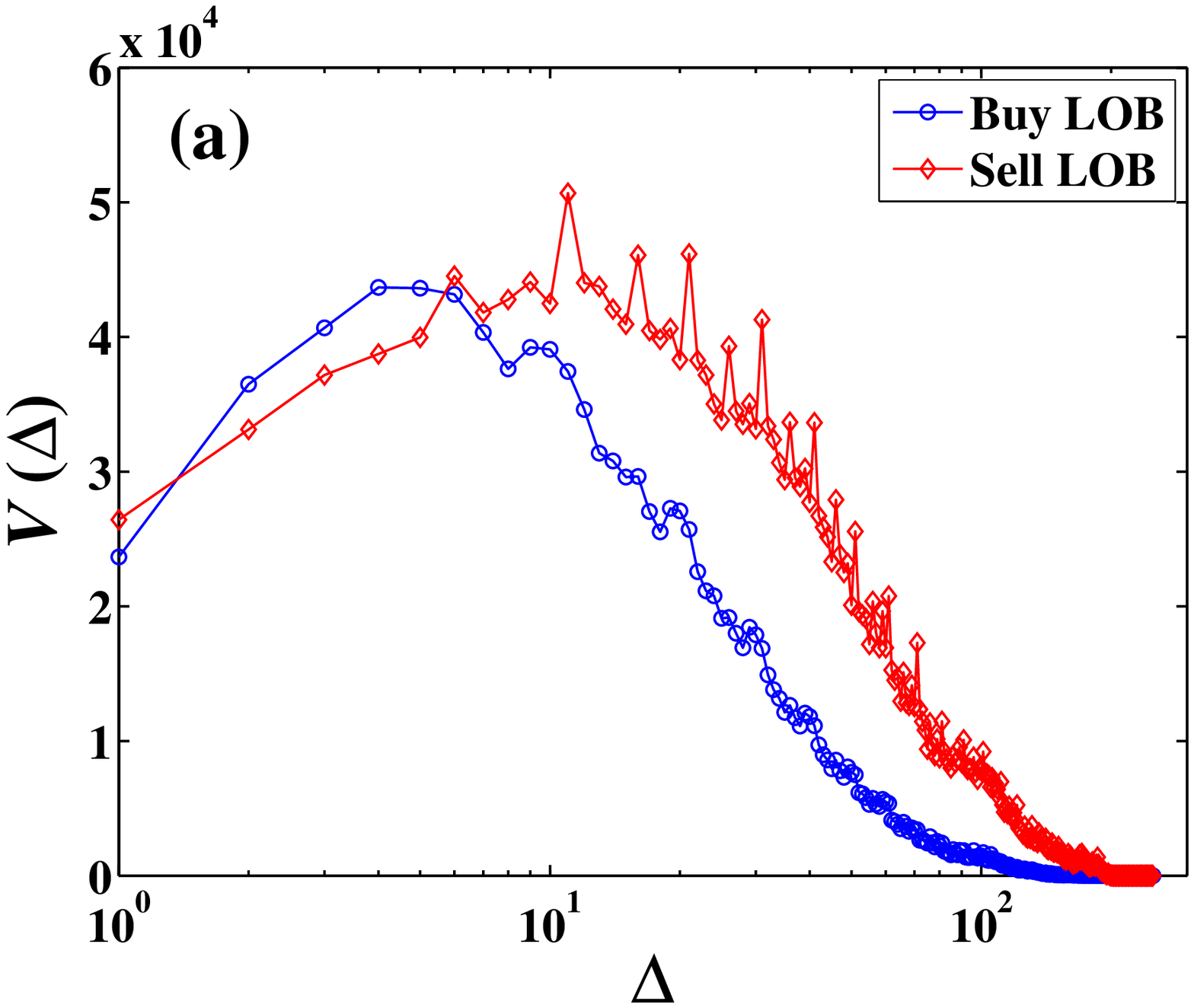}
\includegraphics[width=6.5cm]{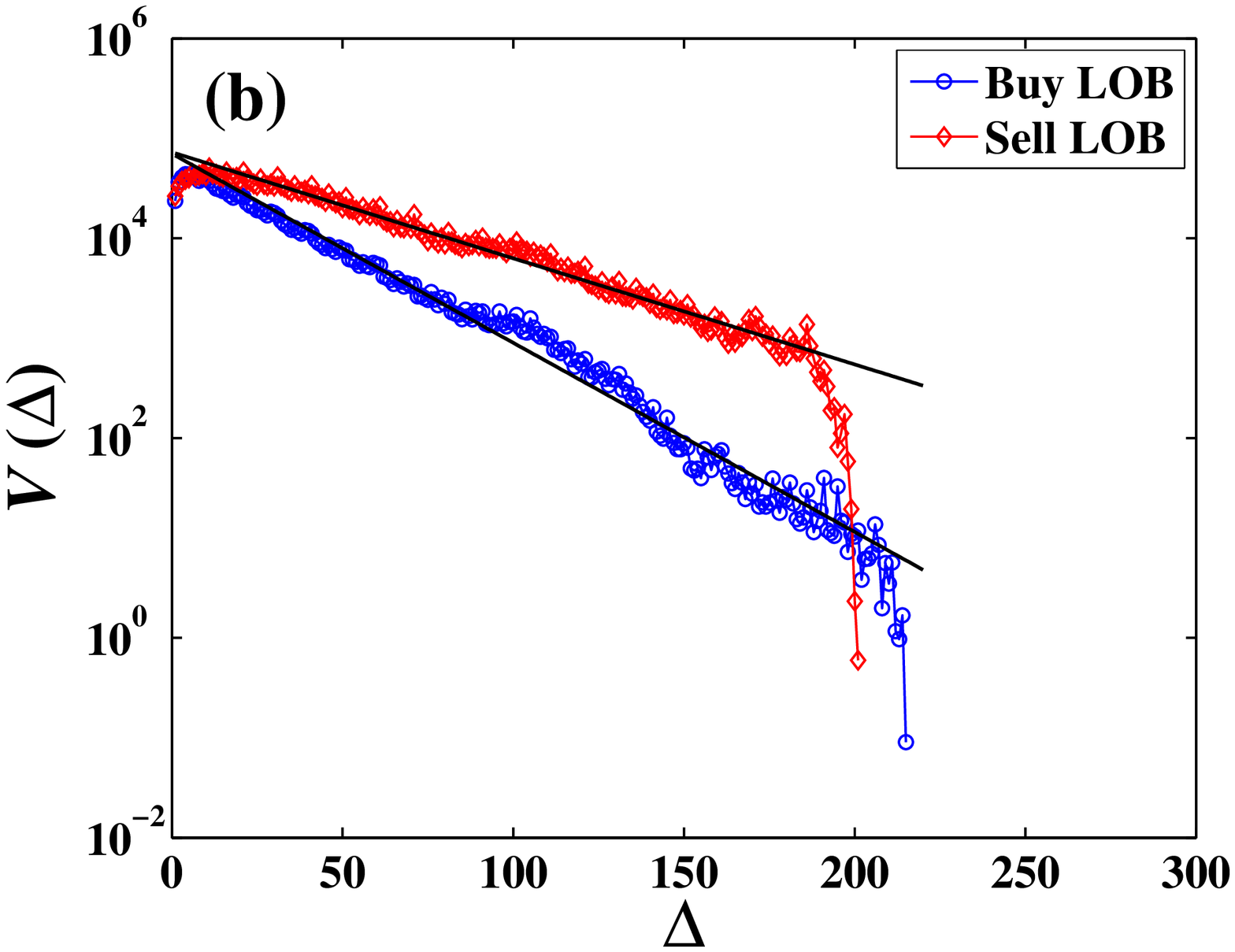}
\caption{\label{Fig:ave_shape} LOB shape $V(\Delta)$ as a function
of relative distance $\Delta$ for buy and sell limit-order books in
log-linear coordinates (a) and linear-log coordinates (b).}
\end{figure}

It is known that traders tend to place their orders on the same best
price
\cite{Bouchaud-Mezard-Potters-2002-QF,Potters-Bouchaud-2003-PA,Maskawa-2007-PA,Mike-Farmer-2008-JEDC,Gu-Chen-Zhou-2007-xxx}.
On the other hand, the orders placed near the same best have a
higher execution probability, and impatient traders are likely to
make a cancelation when these orders are not executed immediately.
It is thus not clear what is the LOB shape under these opposite
forces. Fig.~\ref{Fig:ave_shape} shows the shapes of buy and sell
LOBs.

In Fig.~\ref{Fig:ave_shape}(a), we in general find that the LOB
shape function has a maximum away from the same best ($\Delta=1$)
and is roughly symmetrical to the maximum, which consists with the
result of Bouchaud {\em et al}.
\cite{Bouchaud-Mezard-Potters-2002-QF}. The LOB shapes are
asymmetric between buy orders and sell orders. The LOB shape
$V(\Delta)$ increases when $\Delta\leqslant\Delta_{\max}$ and
decreases afterwards, where $\Delta_{\max}=4$ for buys and
$\Delta_{\max}=11$ for sells. We note that only two (000088 and
000539) of the 23 stocks do not have clear maxima and the values of
$\Delta_{\max}$ vary from stock to stock. In addition, the total
volume of sell orders is greater than that of buy orders, which is
especially visible for large $\Delta$. This phenomenon is also
observed for other stocks except that two stocks (000088 and 000089)
have comparable buy and sell volumes, which is consistent with the
fact that the Chinese stock market in 2003 was in the middle of a
long-lasting bearish antibubble from 2001 to 2005
\cite{Zhou-Sornette-2004a-PA} and more market participators tended
to sell their shares.

There are two more features arise in the empirical LOB shape
function. Although there are more sell limit orders in the book, the
buy LOB is still thinker than sell LOB for small $\Delta$ in
Fig.~\ref{Fig:ave_shape}(a). In 2003, only the information on the
first three visible levels ($\Delta=1$, $n_2$, and $n_3$ such that
the instant LOB shape function $V(n_1)\neq0$, $V(n_2)\neq0$ and
$V(\Delta)=0$ for other relative distances less than $n_3$) were
disposed to traders. We find that, 10 stocks have thicker sell
books, 10 stocks have thicker buy books, and the other three have
comparable book thickness. This observation is very interesting
since the traders faced a very strong illusionary signal that there
were more buy orders while the market was bearish. Another
interesting feature is the presence of periodic peaks at
$\Delta=5n+1$ for $n=0, 1, 2, \cdots$, which are observed in all 23
stocks. The periodic peaks are higher for sell orders than buy
orders. The underlying mechanism of this universal behavior is
unclear, which might be related to the trading strategy of larger
traders or people's irrational preference of some numbers like 5, 10
or their multiples \cite{Dorogovtsev-Mendes-Oliveira-2006-PA}. These
two features call for further investigation, which is however beyond
the scope of this work.

In Fig.~\ref{Fig:ave_shape}(b), we show the shape functions in
linear-log coordinates to study the functional form for large
$\Delta$. The volumes in both buy and sell LOBs decrease
exponentially,
\begin{equation}
 V_{b,s}(\Delta) \sim e^{-\beta_{b,s} \Delta}~.
 \label{Eq:V1}
\end{equation}
Using least-squares fitting method, we obtain that $\beta_b = 0.044
\pm 0.0004$ for buy LOB and $\beta_s = 0.025 \pm 0.0002$ for sell
LOB. The decreasing speed of buy LOB is faster than that of sell
LOB, which means that there is a larger proportion of more
aggressive orders in the buy LOB than in the sell LOB. It seems that
buyers pay more attention to the execution probability, while
sellers consider the return of their investigation more important.
We notice that most of other stocks have similar exponentially
decreasing shapes. In contrast, Bouchaud {\em{et al}}. have found
that the LOB shape tails have power-law behaviors for the three
liquid stocks traded on the Paris Bourse
\cite{Bouchaud-Mezard-Potters-2002-QF}. In addition,
$V_{b,s}(\Delta)$ abruptly plummet to zero at the tail ends, which
is caused by the 10\% price fluctuation limitation compared to the
close price on the previous trading day.

\begin{figure}[htb]
\centering
\includegraphics[width=8cm]{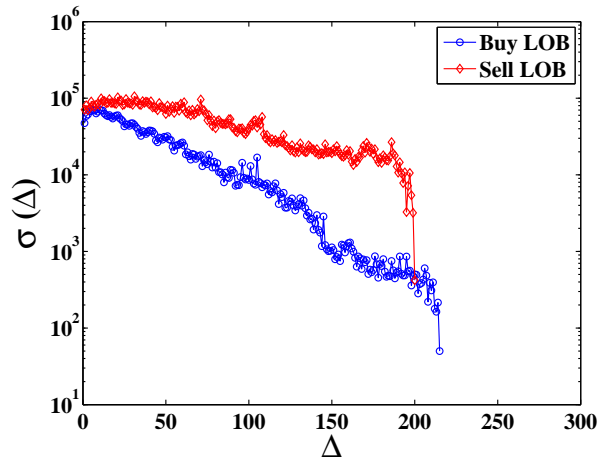}
\caption{\label{Fig:sigma} Plot of the standard deviations
$\sigma(\Delta)$ as a function of the relative distance $\Delta$ for
buy and sell LOBs.}
\end{figure}

We have studied the event-time averaged volume placed at each tick
levels in the LOB. However, the volume may have large fluctuations
and greatly deviate from the mean. It is necessary to analyze the
fluctuations of volumes at each tick levels. Here, we study the
standard deviation $\sigma$ as a function of the relative distance
$\Delta$, that is,
\begin{equation}
\sigma_{b,s}(\Delta) = \sqrt{\langle V_{b,s}(\Delta)^2 \rangle -
\langle V_{b,s}(\Delta) \rangle ^2}~.
 \label{Eq:sigma}
\end{equation}
The standard deviations for buy and sell LOBs are presented in
Fig.~\ref{Fig:sigma}. We find that the functional form of
$\sigma(\Delta)$ is very similar to that of the shape for both buy
and sell LOBs. The standard deviation $\sigma(\Delta)$ increases
with $\Delta$ at the first few levels and then decreases
exponentially. When comparing the buy and LOBs, the sell LOB is
found to be thicker with larger fluctuations.

\section{Statistical properties of volumes at individual tick levels}
\label{sp}

\subsection{Probability distribution}
\label{pdf}

We have analyzed the averaged volume above. Here we focus on the
time averaged volume over a fixed clock time interval $\delta{t}$ at
individual levels
\begin{equation}
v_{b,s}(\Delta,t) =
 \frac{1}{N}\sum_{t = 1}^N{V_{b,s}(\Delta,t_i)}~, \\
\label{Eq:Vt}
\end{equation}
where $t_i$ is the time moments of the $N$ events occur in the
interval $(t-{\delta} t,t]$ and $N$ is a function of $t$ and
$\delta{t}$. We use ${\delta} t = 1$ min to calculate the
time-averaged volume at each price level.

\begin{figure}[htb]
\centering
\includegraphics[width=6.5cm]{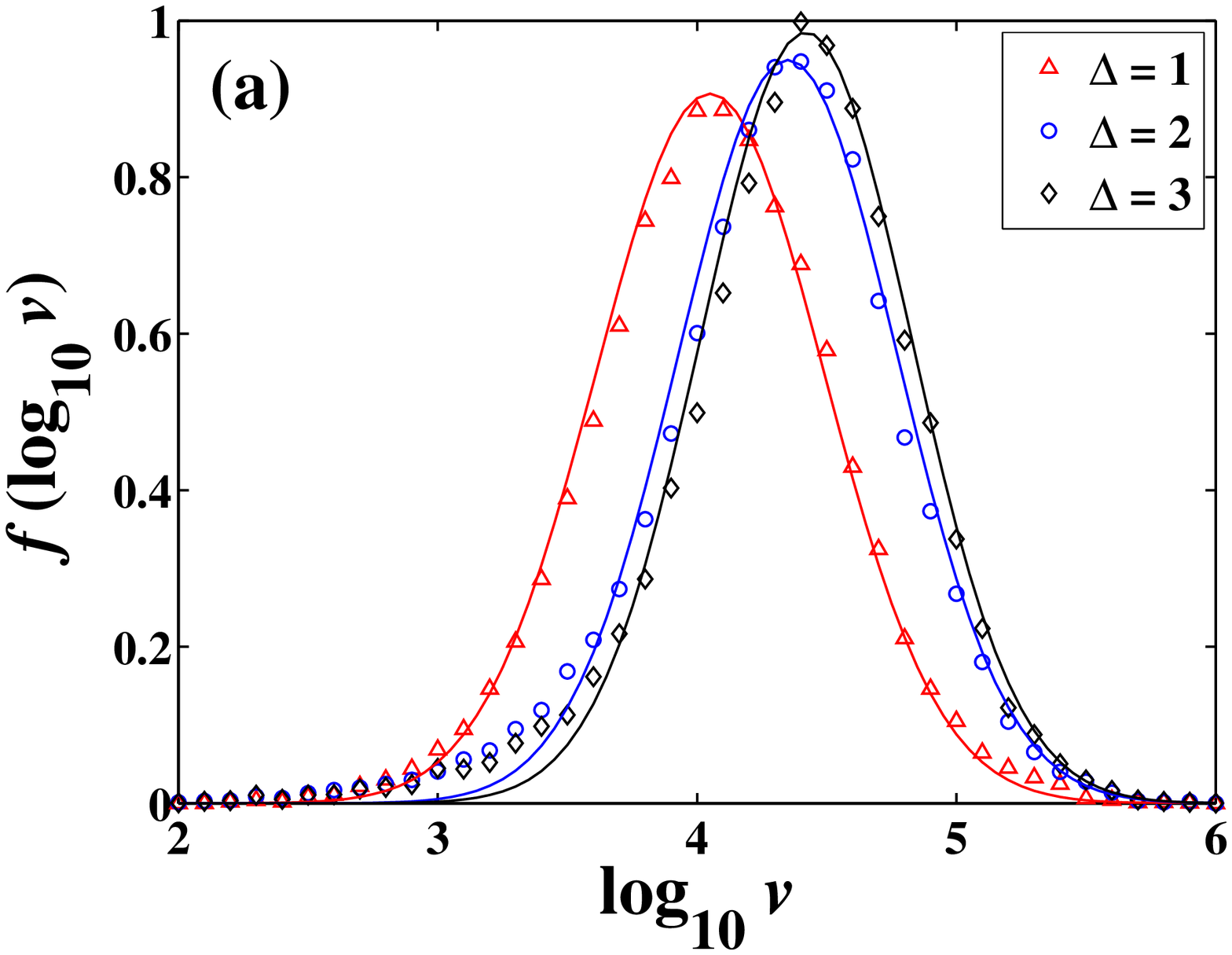}
\includegraphics[width=6.5cm]{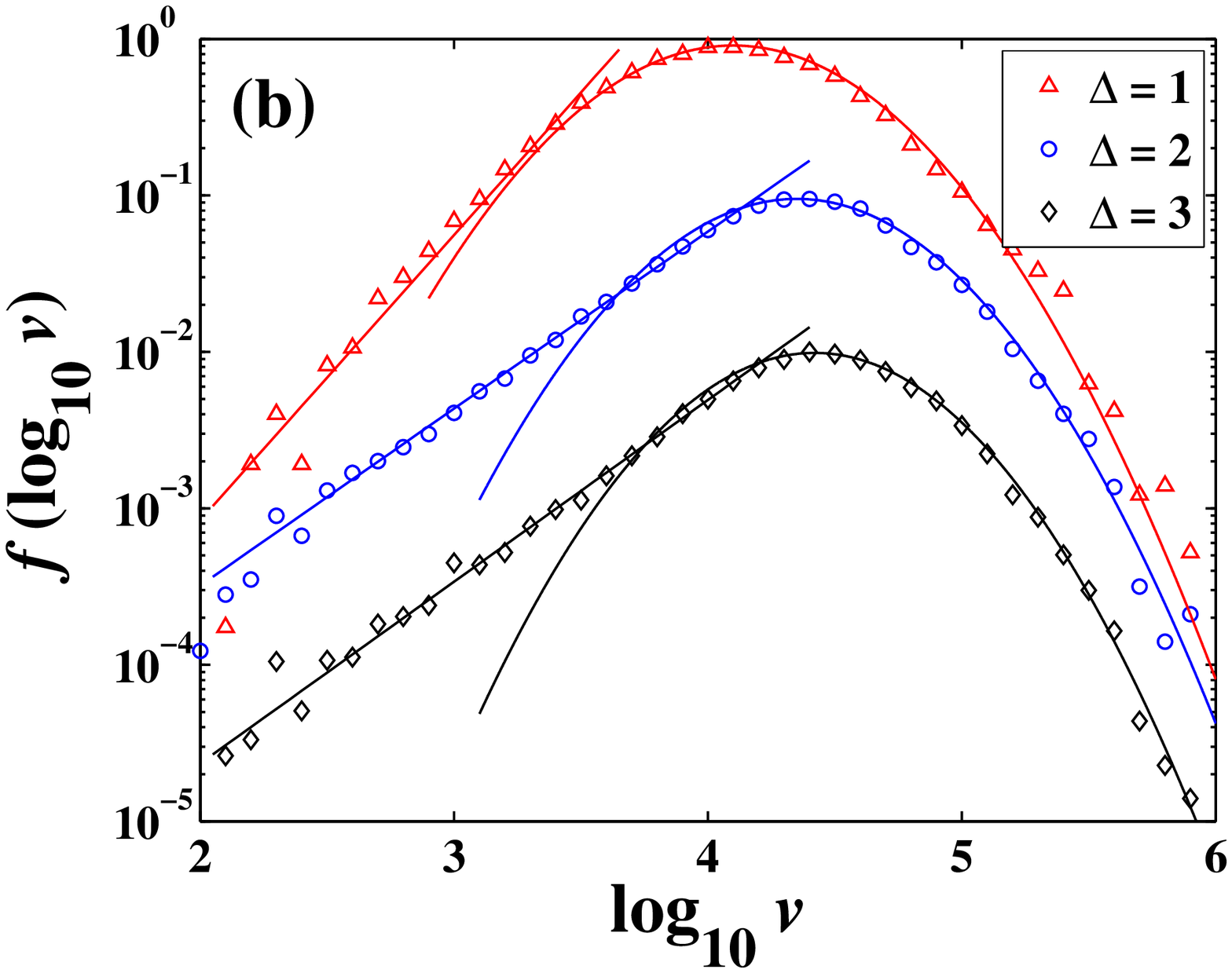}
\caption{\label{Fig:pd_b_123} Probability density functions
$f(\ln{v})$ of 1-min averaged logarithmic volumes at the first three
tick levels on the buy LOB in a linear-linear scale (a) and
linear-log scale (b). The curves corresponding to $\Delta = 2$ and
$\Delta = 3$ in (b) have been vertically translated downward for
clarity. The results are similar on the sell side. }
\end{figure}

Fig.~\ref{Fig:pd_b_123} shows the probability density functions
(PDFs) for $\Delta=1$, 2, and 3. In Fig.~\ref{Fig:pd_b_123} (a), we
find that $\ln{v}$ in general is normally distributed
\begin{equation}
 f(\ln{v}) =
 \frac{1}{\sqrt{2\pi}\sigma}
 \exp\left[-\frac{(\ln{v}-\mu)^2}{2\pi\sigma^2}\right]~,
 \label{Eq:fv}
\end{equation}
that is, $v$ is log-normally distributed with the PDF
being\footnote{Denote $g(y)$ and $h(x)$ the PDFs of $y$ and $x$,
respectively. If $y$ is a function of $x$, we have $g(y)dy =
h(x)dx$. It follows immediately that
$h(x)=g(y){dy}/{dx}={g(\ln{x})}/{x}$.}
\begin{equation}
 p(v) = {f(\ln{v})}/{v}
 \label{Eq:pv}
\end{equation}
This is also different from the Paris Bourse stocks where the
volumes on the best are distributed according to a Gamma
distribution \cite{Bouchaud-Mezard-Potters-2002-QF}. With the
increase of the relative distance $\Delta$, the mean of $\ln{v}$,
$\mu$, increases, which is line with the result in
Fig.~\ref{Fig:ave_shape}. We can also project that $\mu$ decreases
for large $\Delta$. More generally, we find that the 1-min volumes
at other tick levels for different stocks are basically lognormally
distributed.

When $v$ is small, we find that the empirical curves deviate from
the lognormal distribution $f(\ln{v})$. We plot the probability
density functions $f(\ln{v})$ of $\ln{v}$ in a linear-log scale,
which is presented in Fig.~\ref{Fig:pd_b_123} (b). It is clear that
the small volumes $v$ deviate from the corresponding lognormal
distributions and exhibit power-law behaviors
\begin{equation}
 f(\ln{v})\sim v^{\beta_{\Delta}}~~{\rm{or}}~~p(v) \sim
 v^{\beta_{\Delta}-1}~.
 \label{Eq:pv2}
\end{equation}
Using least-squares fitting, we obtain that $\beta_1 = 4.19 \pm
0.09$ ($2.2 < \log_{10}v < 3.5$) for $\Delta = 1$, $\beta_2 = 2.61
\pm 0.03$ ($2.1 < \log_{10}v < 4.2$) for $\Delta = 2$, and $\beta_3
= 2.67 \pm 0.05$ ($2.1 < \log_{10}v < 4.2$) for $\Delta = 3$.

\subsection{Long memory}
 \label{lm}

Temporal dependency can be quantitatively assessed by the
autocorrelation function $C(\ell)$, which describes the average
correlation between two points with time lag $\ell$. Many processes
have the autocorrelation function decaying exponentially ($C(\ell)
\sim e^{-\ell/\ell_0}$ for $\ell \rightarrow \infty$), which means
these processes exhibit short memory with a characteristic timescale
$\ell_0$. On the other hand, when the autocorrelation function is
not integrable, for example, $C(\ell)$ decaying as a power-law
behavior ($C(\ell) \sim \ell^{-\gamma}$), the process has long
memory without any characteristic timescale, which means that the
values in the past have potential predictive power for the future.

\begin{figure}[htb]
\centering
\includegraphics[width=8cm]{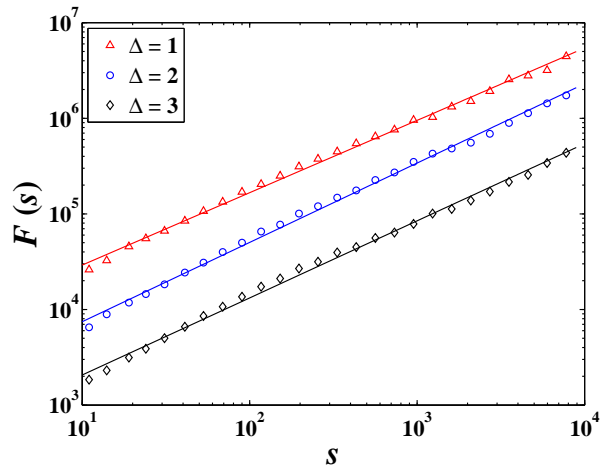}
\caption{\label{Fig:dfa} Plot of the detrended fluctuation functions
$F(\ell)$ of 1-min averaged volumes at the first three tick levels
on the buy limit-order book. The results corresponding to $\Delta =
2$ and $\Delta = 3$ have been vertically translated downwards for
clarity.}
\end{figure}

The property of temporal dependency is equivalently characterized by
the Hurst index $H$, and the relationship between the
autocorrelation exponent ${\gamma}$ (assuming $C(\ell) \sim
\ell^{-\gamma}$) and the Hurst index $H$ can be expressed by
$\gamma=2-2H$
\cite{Kantelhardt-Bunde-Rego-Havlin-Bunde-2001-PA,Maraun-Rust-Timmer-2004-NPG}.
Detrended fluctuation analysis (DFA) is a popular method to estimate
the Hurst index
\cite{Peng-Buldyrev-Havlin-Simons-Stanley-Goldberger-1994-PRE,Hu-Ivanov-Chen-Carpena-Stanley-2001-PRE,Kantelhardt-Bunde-Rego-Havlin-Bunde-2001-PA}.
We perform DFA on the 1-min averaged volumes at the first three tick
levels on the buy LOB. The detrended fluctuation functions $F(\ell)$
are presented in Fig.~\ref{Fig:dfa}. Sound power-law relations are
observed in the three curves and the Hurst indexes are $H_1 = 0.76
\pm 0.01$ for $\Delta = 1$, $H_2 = 0.83 \pm 0.01$ for $\Delta = 2$,
and $H_3 = 0.81 \pm 0.01$ for $\Delta = 3$, respectively. With the
Hurst indexes $H$ significantly larger than 0.5, we argue that the
1-min averaged volumes at the first three tick levels exhibit long
memory. Quantitatively similar results are observed for the sell LOB
and for other stocks. This agrees well with the fact that order
signs have long memory
\cite{Bouchaud-Gefen-Potters-Wyart-2004-QF,Lillo-Farmer-2004-SNDE}.

\section{Conclusion}
 \label{conclusion}

We have investigated the limit-order book shapes of 23 stocks traded
on the Shenzhen Stock Exchange in the whole year 2003. For brevity,
we presented the results of a very liquid stock (Shenzhen
Development Bank Co., LTD, 000001). For most of the stocks, the
averaged shape has a maximum away from the same best and the volumes
in the LOBs decrease exponentially. The LOB shapes are asymmetric
between buy and sell orders and the sell LOB shape relaxes much
slower. The probability density functions of 1-min averaged volumes
at the first three tick levels follow lognormal distributions with a
power-law behavior for small volumes in the left tails. Using
detrended fluctuation analysis, we confirmed that the 1-min averaged
volumes at a fixed tick level on the LOB exhibit long memory. When
compared with the Paris Bourse stocks
\cite{Bouchaud-Mezard-Potters-2002-QF}, we find that the LOB shapes
are qualitative similar but quantitatively different.

Several problems arise that need to be addressed: why the buy LOB is
abnormally thicker for the price levels close to the same best and
why there are relatively large volume on the tick levels of
$\Delta=5n+1$? It is also noteworthy that our results on the
empirical LOB shape functions can be used to develop more realistic
optimal trading strategy for large traders.

\bigskip

{\textbf{Acknowledgments:}}

This work was partly supported by the National Natural Science
Foundation of China (Grant Nos. 70501011 and 70502007), the Fok Ying
Tong Education Foundation (Grant No. 101086), and the Program for
New Century Excellent Talents in University (Grant No.
NCET-07-0288).

\bibliography{E:/Papers/Auxiliary/Bibliography}

\end{document}